\documentclass[%
 preprint,
 amsmath,amssymb,
aip,
apl,
]{revtex4-1}
\usepackage{lipsum}
\usepackage{graphicx}
\usepackage{dcolumn}
\usepackage{bm}

\usepackage{color,soul}
\usepackage{epstopdf}
\AppendGraphicsExtensions{.tif}
\begin{document}


\title{Growth-Temperature Dependence of Conductivity at the LaCrO$_3$/SrTiO$_3$ (001) Interface }
\author{A. Al-Tawhid}
 \affiliation{Department of Physics, North Carolina State University, Raleigh, NC 27695, USA}
 
 \author{J. Frick}
 \affiliation{Department of Physics, North Carolina State University, Raleigh, NC 27695, USA}
 
 \author{D.B. Dougherty}
 \affiliation{Department of Physics, North Carolina State University, Raleigh, NC 27695, USA}
 
\author{D.P. Kumah}
\email{dpkumah@ncsu.edu}
 \affiliation{Department of Physics, North Carolina State University, Raleigh, NC 27695, USA}

\date{\today}

\begin{abstract}
The effect of growth conditions on the structural and electronic properties of the polar/non-polar LaCrO$_3$/SrTiO$_3$ (LCO/STO) interface has been investigated. The interface is either insulating or metallic depending on growth conditions. A high sheet carrier concentration of 2x10$^{16}$ cm$^{-2}$ and mobility of 30,000 cm$^2$/V s is reported for the metallic interfaces, which is similar to the quasi-two dimensional gas at the LaAlO$_{3}$/SrTiO$_{3}$ interface with similar growth conditions.  High-resolution synchrotron X-ray-based structural determination of the atomic-scale structures of both metallic and insulating LCO/STO interfaces show chemical intermixing and an interfacial lattice expansion. Angle resolved photoemission spectroscopy of 2 and 4 uc metallic LCO/STO shows no intensity near the Fermi level indicating that the conducting region is occurring deep enough in the substrate to be inaccessible to photoemission spectroscopy. Post-growth annealing in flowing oxygen causes a reduction in the sheet carrier concentration and mobility for the metallic interface while annealing the insulating interface at high temperatures and low oxygen partial pressures results in metallicity. These results highlight the critical role of defects related to oxygen vacancies on the creation of mobile charge carriers at the LCO/STO heterointerface. 

\end{abstract}

\maketitle

\section{Introduction}
Polar/non-polar perovskite interfaces have attracted considerable interest since the discovery of novel interfacial phenomena not found in the constituent bulk materials\cite{ohtomo2004high, brinkman2007magnetic, wong2010metallicity, mannhart2010oxide}. Of considerable interest is the interface between polar LaAlO$_3$ (LAO) and non-polar SrTiO$_3$(STO), which exhibits  a high mobility quasi-two dimensional electron gas (q-2deg), ferromagnetism and superconductivity, despite both materials being non-magnetic bulk insulators\cite{ohtomo2004high, reyren2007superconducting,brinkman2007magnetic}. Conductivity at the LAO/STO interface has been attributed to an interfacial polarity driven electronic reconstruction,\cite{ohtomo2004high} ionic intermixing\cite{willmott2007structural} and defects related to cation non-stoichiometry\cite{breckenfeld2013effect} and oxygen vacancies.\cite{herranz2007high} Since it's discovery, other complex oxides such as SmTiO$_3$, GdTiO$_3$, LaTiO$_3$, LaGaO$_3$,and LaVO$_3$ have been grown on STO and have been shown to also exhibit interfacial conductivity\cite{perna2010conducting,moetakef2011electrostatic,kornblum2015oxide,marshall2016pseudogaps, he2012metal,ahmadi2018tuning} while some polar/non-polar interfaces such as the LaCrO$_3$(LCO)/STO were found to be insulating\cite{chambers2011band}. An electric field at the LCO/STO interface observed by photoelectron spectroscopy\cite{chambers2011band} is expected to result in an electronic reconstruction resulting in a conducting interface. The absence of conductivity is attributed to Ti outdiffusion which results in a redistribution of charge in the interfacial CrO$_2$ layers.\cite{chambers2011band} However, mobile charge carriers have been reported for LCO/STO superlattices.\cite{comes2017probing} To gain further insight into the intrinsic and extrinsic mechanisms that trigger conduction at these interfaces the effect of growth conditions is required.

In this paper, we investigate the effect of growth temperature on the electrical and structural properties of the LCO/STO interface using a combination of high-resolution synchrotron X-ray diffraction structural measurements, angle resolved photoemission spectroscopy (ARPES) and temperature-dependent transport measurements. Previous reports on the LAO/STO interface have shown that growth temperature can have a considerable effect on the transport properties of the interface\cite{fete2015growth}.  LCO is a bulk insulator with a bandgap of 3.4 eV\cite{arima1993variation}. At room temperature, it has an orthorhombic crystal structure with a pseudo-cubic lattice parameter of 3.88 \AA, leading to a small lattice mismatch of ≈ 0.5\% with STO\cite{HASHIMOTO2000181}. Here, we find that by varying the growth temperature and post-growth treatments, the interface can be tuned from an insulating to a high mobility metallic state. Samples grown at lower temperature (600 $^o$C) are insulating while higher growth temperatures (800 $^o$C) under identical oxygen pressures leads to metallicity. The metallic samples exhibit high mobilities on the order of 10$^4$ cm$^2$/Vs at 2K similar to that of LAO/STO interface under similar growth conditions as well as reduced STO. \cite{kalabukhov2007effect}  There is an expansion of the out-of-plane lattice constant of the STO at the interface in both samples related to the La-Sr inter-diffusion. The tunability of the inter-facial conductivity under post-growth oxidizing and reducing conditions points to defects related to oxygen vacancies as the origin of the interfacial metallicity.\cite{comes2017probing}

\begin{figure}
   \centering
   \includegraphics[width=\linewidth]{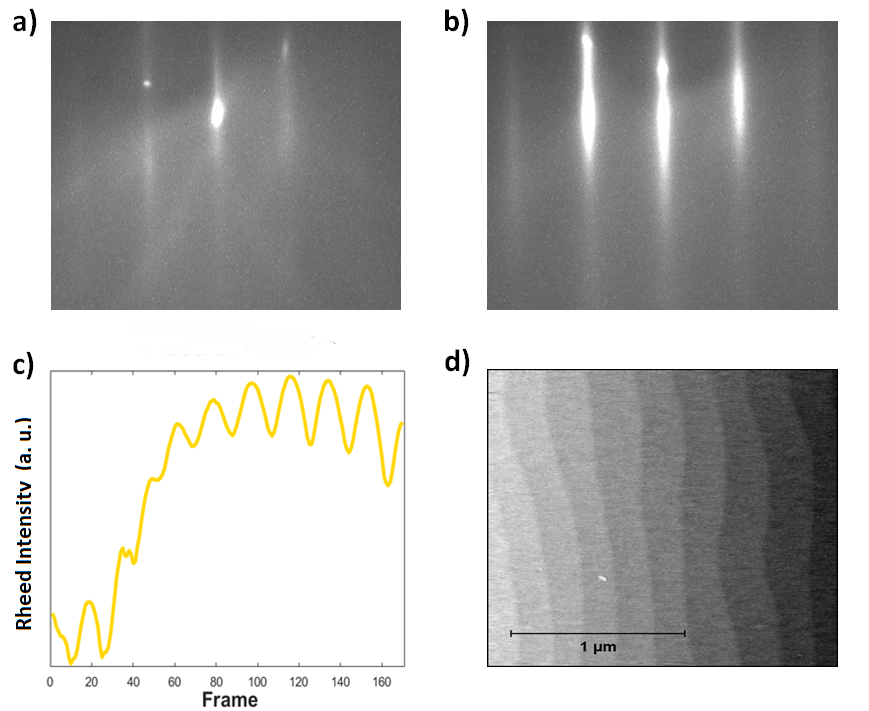}
   \caption{Growth properties of LCO films grown on STO by molecular beam epitaxy. a) RHEED image of bare STO substrate.  b)  RHEED image 10 uc LCO on STO after growth  c) RHEED oscillations for a 10 uc LCO film grown at 800 $^o$C and 1e-7 torr O$_2$.  d) AFM image of a 10 uc LCO. Note that the terrace step structure is preserved with a step height of 4 \AA.}
   \label{fig:1}
 \end{figure}

\section{Experimental Methods}
\subsection{Growth and treatment}
LCO films with a nominal thickness of 10 unit cells (uc) were grown on (001) oriented STO substrates using oxygen assisted molecular beam epitaxy (MBE). Prior to growth, the STO substrates were treated with buffered HF and annealed in flowing oxygen at 1000 $^o$C in a tube furnace for 2 hours to create a Ti terminated surface\cite{kawasaki1994atomic}. Atomic force microscope (AFM) images of the substrate after treatment show that the terrace step structure is present. The LCO films were grown at 1x$10^{-7}$ Torr oxygen partial pressure with the base pressure of the MBE chamber being 2x$10^{-10}$ Torr. The flux of the La and Cr sources were calibrated before growth with a quartz crystal microbalance. \textit{In-situ} reflection high energy electron diffraction (RHEED) was used to monitor the growth process. Images of the RHEED pattern of a bare STO and a 10 uc LCO are shown in Figure 1(a) and 1(b) respectively. RHEED oscillations of a 10 uc LCO in Figure 1(c) indicates that layer-by-layer growth is achieved and the number of layers deposited is determined by the number of RHEED oscillations. AFM images after growth of the LCO (Figure 1(d)) shows that the terrace step structure of the STO substrate is preserved confirming epitaxial growth.  
 The samples were grown in two temperature regimes; a high temperature regime(800 $^o$C) (henceforth referred to as HT) and a low temperature regime(600 $^o$C),(henceforth referred to as LT) with two samples grown at each temperature. One sample from each regime was selected for post-growth treatment. One HT sample was annealed in flowing oxygen at 300 $^o$C for 1.5 hr in a tube furnace and a LT sample was annealed at 800 $^o$C in 1x$10^{-7}$ Torr oxygen partial pressure for 30 minutes in the MBE chamber. 

\begin{figure*}[ht]
\includegraphics[scale=0.36]{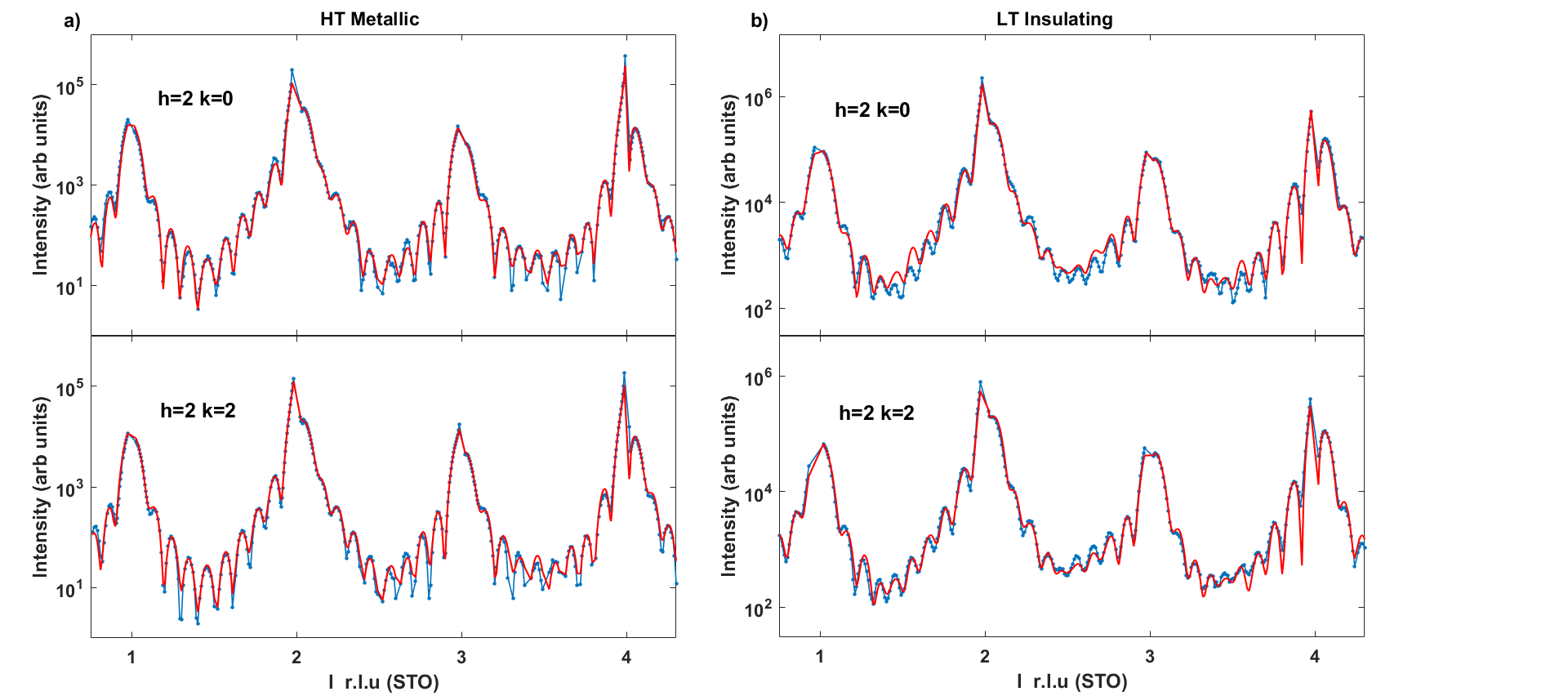} 
\caption{ Representative measured (blue circles)  and calculated (red curves) crystal truncation rods (CTRs) for (a) HT as-grown 10 uc LCO/STO heterostructure grown at 800 $^o$C and  (b) an a LT as-grown 10 uc LCO/STO heterostructure grown at 600 $^o$C. The h,k and l values are in units of the bulk cubic SrTiO$_3$ reciprocal lattice vectors.   }
\label{fig:CTR}
\end{figure*}

\subsection{Transport/Hall Measurements}

Temperature-dependent Hall and resistivity measurements were performed using a 4-point Van der Pauw configuration in the range of 2 to 300 K. Gold contacts were deposited at the corners of the samples using a shadow mask. All transport measurement were performed using a Quantum Design physical property measurement system (PPMS). The HT as grown samples were found to be conducting at room temperature with sheet resistivities  on the order of 100$\Omega /\square$ while the LT as grown samples were highly insulating ($>$100 M$\Omega/\square$).

\subsection{Synchrotron X-ray Diffraction}
To determine the differences in the atomic-scale structures of the as-grown HT-metallic and LT-insulating nominally 10 uc LCO films, synchrotron X-ray diffraction measurements were carried out at the 33ID beamline at the Advanced Photon Source. Crystal truncation rods (CTR) were measured along the STO bulk crystallographic directions with an incident X-ray energy of 15.5 eV. The layer-resolved structure and composition of the films were obtained from fitting the CTR data with a genetic based algorithm, GenX\cite{bjorck2007genx}. The parameters of the fit included the out-of- plane-lattice constant of each layer,  Debye-Waller factors for each element and the fractional occupation of the surface LCO layer to account for surface roughness. Additionally, the chemical composition of the first five layers of the film and substrate at the film/substrate interface were optimized to allow for intermixing. Each element was also allowed to displace in (001) direction and the oxygen octahedral were allowed to rotate and tilt.\cite{koohfar2017structural} The fit was then compared to the data iteratively to minimize the error between them. Structural convergence was achieved by minimizing the crystallographic R1 error function. We obtain excellent fits with figures of merit (FoM) values below .08 for each fit. Representative measured CTRs and their associated fits for the HT and LT as grown samples are shown in Figure 2(a) and 2(b) respectively.

\subsection{Angle Resolved Photoemission Spectroscopy}

Angle-resolved photoemission spectroscopy (ARPES) was performed at room temperature (~ 27 $^o$C ) in ultrahigh vacuum (UHV) using both a He I (21.2 eV) plasma light source and a pulsed laser source ( 6.2 eV from the fourth harmonic of a Ti-sapphire laser).  Photoelectrons were analyzed using a 150 mm mean radius hemispherical analyzer (Specs Phoibos 150) with 2D detector.  All samples were transported from the growth chamber through ambient air to be inserted in the ARPES chamber.  They were subsequently annealed in vacuum at 300$^o$C  for 30 minutes and allowed to cool back to room temperature prior to data collection.  The binding energy scale on all samples was corrected for charging effects in order to correspond with K. Maiti et. al.\cite{maiti1996electronic}

\section{Results and Discussion }

\begin{figure}[ht]
\centering
\includegraphics[scale=0.75]{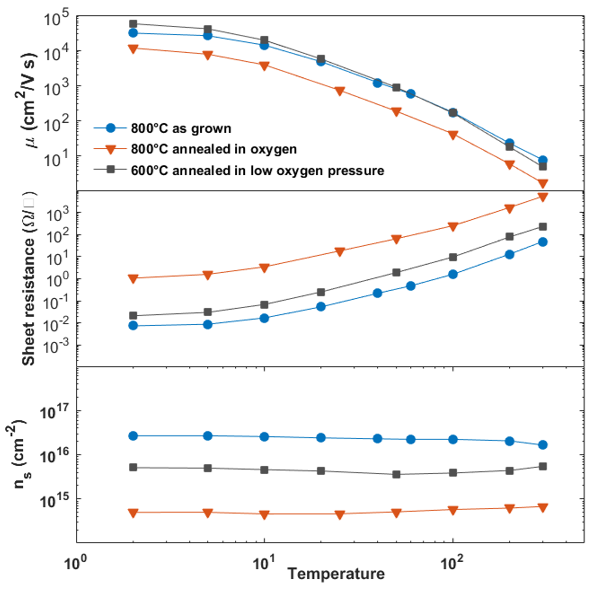} 
\caption{ Comparison of mobility, 4pt sheet resistance and sheet carrier density as a function of growth temperature for i) 10 uc LCO/STO grown at 800 $^o$C without a post-growth anneal,ii) 800 $^o$C growth followed by an anneal in 1 atm 99.99\% O$_2$ at 300 $^o$C, and iii) 600 $^o$C growth followed by an anneal at 800 $^o$C in 1x10$^{-7}$ Torr O$_2$.  }
\label{fig:RVT_anneal}
\end{figure} 
Figure 3 shows the temperature-dependent mobility ($\mu$), sheet resistance and sheet carrier density (n$_s$) for the metallic LCO/STO samples determined from the Hall measurements. The transport properties are compared for the as-grown HT sample, the flowing oxygen-annealed HT sample, and the LT sample annealed at 800 $^o$C in 1x$10^{-7}$ Torr O$_2$.  For all 3 metallic samples, the electron mobility increases as the temperature is reduced while the sheet carrier concentration remains roughly constant. The HT as-grown sample shows a mobility of 30,000 cm$^2$/Vs and a sheet carrier density of 2x10$^{16}$ cm$^{-2}$ at 2 K which is similar to previous reports on the electrical properties of the 2-deg at the LAO/STO interface grown under low oxygen growth pressures and metallic reduced STO substrates\cite{kalabukhov2007effect}. Annealing the HT sample in 1 atm O$_2$ at 300 $^o$C for 1.5 hrs  reduces the sheet carrier concentrations by more than an order of magnitude. 

In contrast to the metallic as-grown HT sample, the as-grown LT sample shows no measurable conductivity at room temperature, hence, we estimate its sheet resistance to be greater than 100 M$\Omega/\square$. Annealing the insulating LT sample under the growth conditions of our metallic sample (800 $^o$ C, 1x$10^{-7}$ Torr oxygen) causes it to become metallic  with a mobility of 10$^4$ cm$^2$/Vs and a sheet carrier concentration of 5x10$^{14}$ cm$^{-2}$ at 2 K. 

To determine the effect of the growth conditions on the structural properties of the LCO/STO interface, the layer-resolved atomic-scale structures were determined from fits to CTRs measured by synchrotron X-ray diffraction. The structural properties of as-grown HT and LT samples are compared in Figure 4. Figure 4(a) shows the A-site composition along the growth direction for the LT and HT samples.  The chemical profile across the interface shows that significant La/Sr inter-diffusion occurs  for both insulating and metallic samples in agreement with previous electron microscopy and Rutherford backscattering measurements which has been suggested to be driven by strain relaxation.\cite{colby2013cation}

For both the HT and LT samples, a tail of Sr outdiffusion into the LCO film extends about 4-5 layers away from the interface (layers 1-5 of Figure 4(a)) decaying from 25\% in the first interfacial LCO layer. Due to the comparable atomic scattering factors of Cr (24 electrons) and Ti (22 electrons) at 15.5 keV, the uncertainty associated with the B-site composition profile are too large to draw relevant conclusions, however, previous high-resolution electron microscopy measurements show Cr-Ti inter-diffusion does occur at the LCO/STO interface.\cite{qiao2011lacro3,colby2013cation}

The layer-resolved lattice constants shown in Figure 4(b) are determined from the spacing of the A-site cations along the growth direction. We observe in both samples an expansion at the interface followed by a linear decrease towards the film surface. The expansion in the STO (layers -1 and 0) is related to the diffusion of La into the STO substrate.\cite{koohfar2017structural,willmott2007structural} La substituting for Sr would donate an electron and alter the valence state of Ti from +4 to +3 increasing its ionic radius. For the LCO film, the expected out-of-plane lattice constant due to the epitaxial strain from the STO substrate is 3.845 \AA. The larger lattice constant in both the HT-metallic and the LT-insulating LCO film may be related to La-Sr and Cr-Ti interdiffusion and/or oxygen vacancies in the LCO. 

\begin{figure}[ht]
\centering
\includegraphics[scale=0.65]
{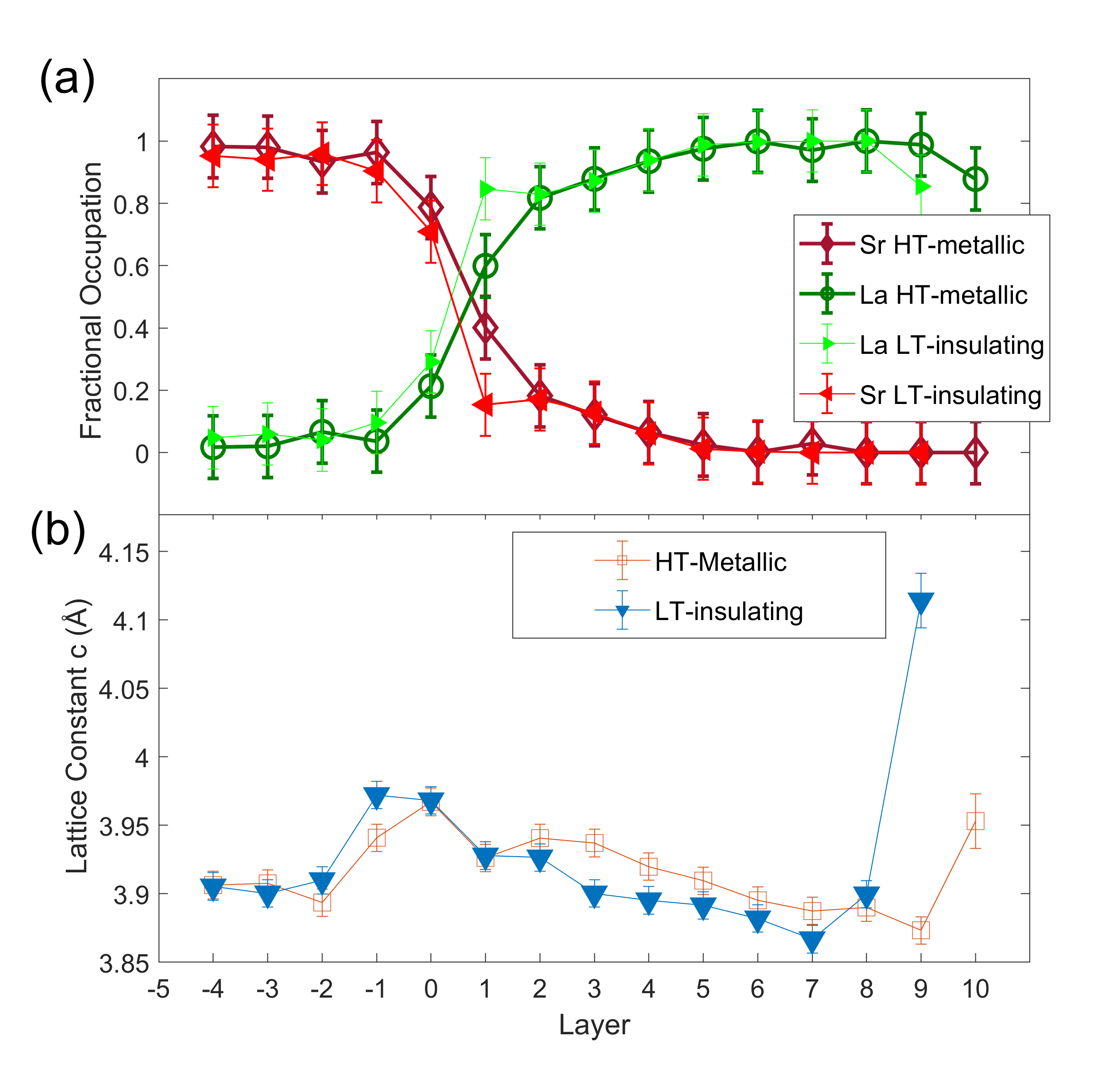}
\caption{(Upper panel). Layer-resolved  Sr/La occupation for for metallic as 800 $^o$C and insulating 600 $^o$C as grown LCO samples. (Lower panel). Layer resolved out of place lattice constant for metallic as 800 $^o$C and insulating 600 $^o$C as grown LCO samples. Layer 0 indicates the top of the STO substrate.}
\label{fig:lattice_constant2}
\end{figure}

Doping due to cation intermixing can be ruled out as the main source of conductivity since it is observed in both the insulating and metallic samples in Figure 4. Hole doping of LCO with Sr can induce a insulator-metal transition for Sr dopants above 65\%.\cite{zhang2015hole} However, the Sr content observed for both films in Figure 4(a) is insufficient to account for the metallicity. Additionally, the Hall coefficient for the metallic samples is negative indicative of majority electron carriers. 

The high carrier concentrations (10$^{14}$-10$^{16}$ cm$^{-2}$) observed for the metallic samples  and the correlation between the growth temperature and post-growth oxygen annealing conditions on metallicity point to oxygen vacancies in the STO as the dominant source of free carriers in the LCO/STO system.\cite{ohtomo2004high, kalabukhov2007effect,plumb2014mixed}

\begin{figure}[ht]
\centering
\includegraphics[scale=0.5]
{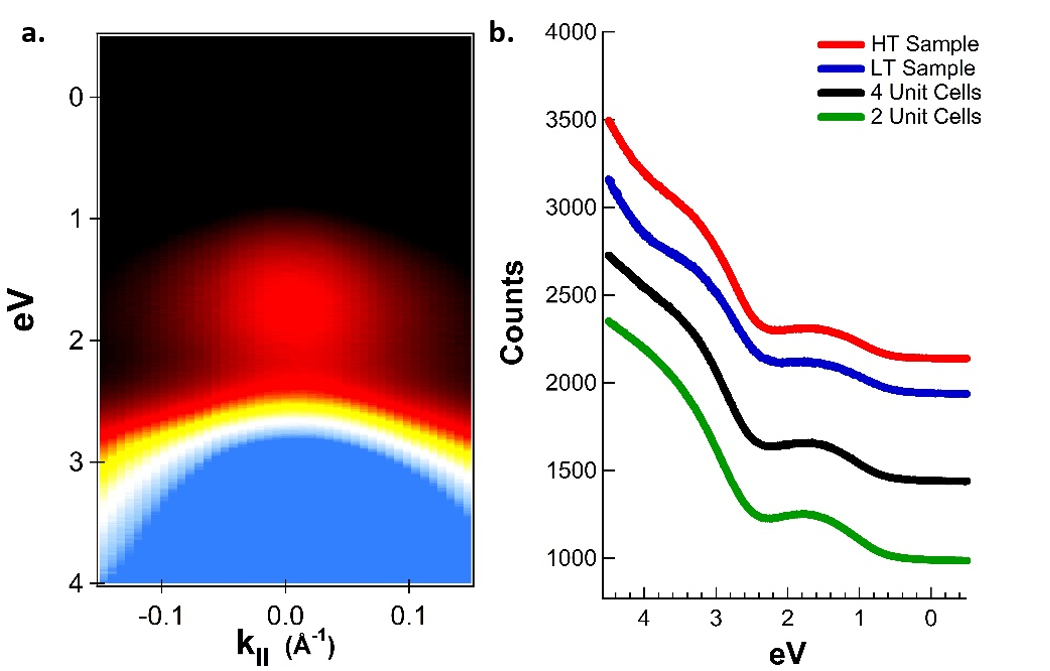}
\caption{(a) Angled-resolved photoemission spectra of 4 unit cell HT LaCrO$_3$ thin film;  (b) Line cuts at k = 0 of a thick (bulk-like) HT LaCrO$_3$ sample (red), thick LT sample (blue), 4 unit cell thick HT sample (black), and 2 unit cell HT sample(green).}
\label{fig:ARPES}
\end{figure}

Angle resolved photoelectron spectroscopy measurements were carried out to elucidate the electronic structure near the surface of the LaCrO$_3$ films.  Figure \ref{fig:ARPES}(a) shows ARPES measurements of a 4 unit cell LaCrO$_3$ film grown under high temperature conditions that was verified by resistivity measurements to show metallic behavior.  Between 1 eV and 2 eV we find a state assigned predominantly to the Cr 3d valence orbitals while the more deeply bound state (2.5 eV $–$ 4.0 eV) comes predominantly from O 2p orbitals\cite{maiti1996electronic} .  The data shown in Figure \ref{fig:ARPES}(a) and \ref{fig:ARPES}(b) are representative of all films studied under different thicknesses, different photon energies, and different substrate temperatures during growth.  Despite these numerous different conditions, all ARPES measurements showed precisely the features seen in Figure \ref{fig:ARPES}(a).
No photo-electron intensity near the Fermi level (0.0 eV) is observed for any of the LCO films considered (Figure \ref{fig:ARPES}(b)).  
The results shown in Figure \ref{fig:ARPES} correspond to 21.2 eV photon energies, which result in photoelectrons with very short escape depths, and thus sensitive mostly to the surface of the film.  However, we performed the same measurements with 6.2 eV photons (not shown here) that are expected to probe significantly more deeply into the bulk of the films and did not find any evidence of a metallic state near the Fermi level for any films.  By extending our measurements down to 2 unit cell films (lowest spectrum in Figure \ref{fig:ARPES}(b)) we confirm that the metallic transport behavior is deeply buried enough below the LCO surface to be inaccessible to photo-electron spectroscopy.

\section{Conclusion}
In summary, we have shown that growth temperature plays a critical role in the creation of mobile charge carriers in LCO/STO(001) heterostructures. Metallic LCO/STO samples exhibit carrier mobilities exceeding 10$^4$ cm$^2$/Vs comparable to the mobility observed at the polar/non-polar LAO/STO interface. Structurally, both insulating and metallic LCO/STO samples exhibit significant La/Sr intermixing at the film-substrate interface and a lattice expansion along the growth direction. Metallicity is correlated with the formation of oxygen vacancies which are tunable by the film growth temperature and post-growth annealing. These results shed light on the critical role of growth parameters on the interfacial properties of perovskite oxide heterostructures. Further work is required to investigate the depth profile of the free carriers as well as determining the effect of the proximity of anti-ferromagnetic LCO\cite{weinberg1961electron,ueda1998ferromagnetism, zhou2011magnetic} on  superconductivity\cite{reyren2007superconducting}  at polar oxide/STO interfaces.

\section{Acknowledgements}
This material is based upon work supported by the National Science Foundation under Grant No. NSF DMR1751455. Photoelectron spectroscopy experiments were supported by US Department of Energy, Office of Science, Basic Energy Sciences under Award No. DE-SC0010324. Use of the Advanced Photon Source was supported by the U. S. Department of Energy, Office of Science, Office of Basic Energy Sciences, under Contract No. DE-AC02-06CH11357.


\begin{thebibliography}{31}%
\makeatletter
\providecommand \@ifxundefined [1]{%
 \@ifx{#1\undefined}
}%
\providecommand \@ifnum [1]{%
 \ifnum #1\expandafter \@firstoftwo
 \else \expandafter \@secondoftwo
 \fi
}%
\providecommand \@ifx [1]{%
 \ifx #1\expandafter \@firstoftwo
 \else \expandafter \@secondoftwo
 \fi
}%
\providecommand \natexlab [1]{#1}%
\providecommand \enquote  [1]{``#1''}%
\providecommand \bibnamefont  [1]{#1}%
\providecommand \bibfnamefont [1]{#1}%
\providecommand \citenamefont [1]{#1}%
\providecommand \href@noop [0]{\@secondoftwo}%
\providecommand \href [0]{\begingroup \@sanitize@url \@href}%
\providecommand \@href[1]{\@@startlink{#1}\@@href}%
\providecommand \@@href[1]{\endgroup#1\@@endlink}%
\providecommand \@sanitize@url [0]{\catcode `\\12\catcode `\$12\catcode
  `\&12\catcode `\#12\catcode `\^12\catcode `\_12\catcode `\%12\relax}%
\providecommand \@@startlink[1]{}%
\providecommand \@@endlink[0]{}%
\providecommand \url  [0]{\begingroup\@sanitize@url \@url }%
\providecommand \@url [1]{\endgroup\@href {#1}{\urlprefix }}%
\providecommand \urlprefix  [0]{URL }%
\providecommand \Eprint [0]{\href }%
\providecommand \doibase [0]{http://dx.doi.org/}%
\providecommand \selectlanguage [0]{\@gobble}%
\providecommand \bibinfo  [0]{\@secondoftwo}%
\providecommand \bibfield  [0]{\@secondoftwo}%
\providecommand \translation [1]{[#1]}%
\providecommand \BibitemOpen [0]{}%
\providecommand \bibitemStop [0]{}%
\providecommand \bibitemNoStop [0]{.\EOS\space}%
\providecommand \EOS [0]{\spacefactor3000\relax}%
\providecommand \BibitemShut  [1]{\csname bibitem#1\endcsname}%
\let\auto@bib@innerbib\@empty
\bibitem [{\citenamefont {Ohtomo}\ and\ \citenamefont
  {Hwang}(2004)}]{ohtomo2004high}%
  \BibitemOpen
  \bibfield  {author} {\bibinfo {author} {\bibfnamefont {A.}~\bibnamefont
  {Ohtomo}}\ and\ \bibinfo {author} {\bibfnamefont {H.}~\bibnamefont {Hwang}},\
  }\href@noop {} {\bibfield  {journal} {\bibinfo  {journal} {Nature}\ }\textbf
  {\bibinfo {volume} {427}},\ \bibinfo {pages} {423} (\bibinfo {year}
  {2004})}\BibitemShut {NoStop}%
\bibitem [{\citenamefont {Brinkman}\ \emph {et~al.}(2007)\citenamefont
  {Brinkman}, \citenamefont {Huijben}, \citenamefont {Van~Z.}, \citenamefont
  {Huijben}, \citenamefont {Zeitler}, \citenamefont {Maan}, \citenamefont
  {van~der Wiel}, \citenamefont {Rijnders}, \citenamefont {Blank},\ and\
  \citenamefont {Hilgenkamp}}]{brinkman2007magnetic}%
  \BibitemOpen
  \bibfield  {author} {\bibinfo {author} {\bibfnamefont {A.}~\bibnamefont
  {Brinkman}}, \bibinfo {author} {\bibfnamefont {M.}~\bibnamefont {Huijben}},
  \bibinfo {author} {\bibfnamefont {M.}~\bibnamefont {Van~Z.}}, \bibinfo
  {author} {\bibfnamefont {J.}~\bibnamefont {Huijben}}, \bibinfo {author}
  {\bibfnamefont {U.}~\bibnamefont {Zeitler}}, \bibinfo {author} {\bibfnamefont
  {J.}~\bibnamefont {Maan}}, \bibinfo {author} {\bibfnamefont {W.~G.}\
  \bibnamefont {van~der Wiel}}, \bibinfo {author} {\bibfnamefont
  {G.}~\bibnamefont {Rijnders}}, \bibinfo {author} {\bibfnamefont {D.~H.}\
  \bibnamefont {Blank}}, \ and\ \bibinfo {author} {\bibfnamefont
  {H.}~\bibnamefont {Hilgenkamp}},\ }\href@noop {} {\bibfield  {journal}
  {\bibinfo  {journal} {Nat. Mat.}\ }\textbf {\bibinfo {volume} {6}},\ \bibinfo
  {pages} {493} (\bibinfo {year} {2007})}\BibitemShut {NoStop}%
\bibitem [{\citenamefont {Wong}\ \emph {et~al.}(2010)\citenamefont {Wong},
  \citenamefont {Baek}, \citenamefont {Chopdekar}, \citenamefont {Mehta},
  \citenamefont {Jang}, \citenamefont {Eom},\ and\ \citenamefont
  {Suzuki}}]{wong2010metallicity}%
  \BibitemOpen
  \bibfield  {author} {\bibinfo {author} {\bibfnamefont {F.~J.}\ \bibnamefont
  {Wong}}, \bibinfo {author} {\bibfnamefont {S.-H.}\ \bibnamefont {Baek}},
  \bibinfo {author} {\bibfnamefont {R.~V.}\ \bibnamefont {Chopdekar}}, \bibinfo
  {author} {\bibfnamefont {V.~V.}\ \bibnamefont {Mehta}}, \bibinfo {author}
  {\bibfnamefont {H.-W.}\ \bibnamefont {Jang}}, \bibinfo {author}
  {\bibfnamefont {C.-B.}\ \bibnamefont {Eom}}, \ and\ \bibinfo {author}
  {\bibfnamefont {Y.}~\bibnamefont {Suzuki}},\ }\href@noop {} {\bibfield
  {journal} {\bibinfo  {journal} {Phys. Rev. B}\ }\textbf {\bibinfo {volume}
  {81}},\ \bibinfo {pages} {161101} (\bibinfo {year} {2010})}\BibitemShut
  {NoStop}%
\bibitem [{\citenamefont {Mannhart}\ and\ \citenamefont
  {Schlom}(2010)}]{mannhart2010oxide}%
  \BibitemOpen
  \bibfield  {author} {\bibinfo {author} {\bibfnamefont {J.}~\bibnamefont
  {Mannhart}}\ and\ \bibinfo {author} {\bibfnamefont {D.}~\bibnamefont
  {Schlom}},\ }\href@noop {} {\bibfield  {journal} {\bibinfo  {journal}
  {Science}\ }\textbf {\bibinfo {volume} {327}},\ \bibinfo {pages} {1607}
  (\bibinfo {year} {2010})}\BibitemShut {NoStop}%
\bibitem [{\citenamefont {Reyren}\ \emph {et~al.}(2007)\citenamefont {Reyren},
  \citenamefont {Thiel}, \citenamefont {Caviglia}, \citenamefont {Kourkoutis},
  \citenamefont {Hammerl}, \citenamefont {Richter}, \citenamefont {Schneider},
  \citenamefont {Kopp}, \citenamefont {R{\"u}etschi}, \citenamefont {Jaccard},\
  and\ \citenamefont {M.}}]{reyren2007superconducting}%
  \BibitemOpen
  \bibfield  {author} {\bibinfo {author} {\bibfnamefont {N.}~\bibnamefont
  {Reyren}}, \bibinfo {author} {\bibfnamefont {S.}~\bibnamefont {Thiel}},
  \bibinfo {author} {\bibfnamefont {A.}~\bibnamefont {Caviglia}}, \bibinfo
  {author} {\bibfnamefont {L.~F.}\ \bibnamefont {Kourkoutis}}, \bibinfo
  {author} {\bibfnamefont {G.}~\bibnamefont {Hammerl}}, \bibinfo {author}
  {\bibfnamefont {C.}~\bibnamefont {Richter}}, \bibinfo {author} {\bibfnamefont
  {C.}~\bibnamefont {Schneider}}, \bibinfo {author} {\bibfnamefont
  {T.}~\bibnamefont {Kopp}}, \bibinfo {author} {\bibfnamefont {A.}~\bibnamefont
  {R{\"u}etschi}}, \bibinfo {author} {\bibfnamefont {D.}~\bibnamefont
  {Jaccard}}, \ and\ \bibinfo {author} {\bibfnamefont {G.}~\bibnamefont {M.}},\
  }\href@noop {} {\bibfield  {journal} {\bibinfo  {journal} {Science}\ }\textbf
  {\bibinfo {volume} {317}},\ \bibinfo {pages} {1196} (\bibinfo {year}
  {2007})}\BibitemShut {NoStop}%
\bibitem [{\citenamefont {Willmott}\ \emph {et~al.}(2007)\citenamefont
  {Willmott}, \citenamefont {Pauli}, \citenamefont {Herger}, \citenamefont
  {Schlep{\"u}tz}, \citenamefont {Martoccia}, \citenamefont {Patterson},
  \citenamefont {Delley}, \citenamefont {Clarke}, \citenamefont {Kumah},
  \citenamefont {Cionca} \emph {et~al.}}]{willmott2007structural}%
  \BibitemOpen
  \bibfield  {author} {\bibinfo {author} {\bibfnamefont {P.}~\bibnamefont
  {Willmott}}, \bibinfo {author} {\bibfnamefont {S.}~\bibnamefont {Pauli}},
  \bibinfo {author} {\bibfnamefont {R.}~\bibnamefont {Herger}}, \bibinfo
  {author} {\bibfnamefont {C.}~\bibnamefont {Schlep{\"u}tz}}, \bibinfo {author}
  {\bibfnamefont {D.}~\bibnamefont {Martoccia}}, \bibinfo {author}
  {\bibfnamefont {B.}~\bibnamefont {Patterson}}, \bibinfo {author}
  {\bibfnamefont {B.}~\bibnamefont {Delley}}, \bibinfo {author} {\bibfnamefont
  {R.}~\bibnamefont {Clarke}}, \bibinfo {author} {\bibfnamefont
  {D.}~\bibnamefont {Kumah}}, \bibinfo {author} {\bibfnamefont
  {C.}~\bibnamefont {Cionca}},  \emph {et~al.},\ }\href@noop {} {\bibfield
  {journal} {\bibinfo  {journal} {Phys. Rev. Lett.}\ }\textbf {\bibinfo
  {volume} {99}},\ \bibinfo {pages} {155502} (\bibinfo {year}
  {2007})}\BibitemShut {NoStop}%
\bibitem [{\citenamefont {Breckenfeld}\ \emph {et~al.}(2013)\citenamefont
  {Breckenfeld}, \citenamefont {Bronn}, \citenamefont {Karthik}, \citenamefont
  {Damodaran}, \citenamefont {Lee}, \citenamefont {Mason},\ and\ \citenamefont
  {Martin}}]{breckenfeld2013effect}%
  \BibitemOpen
  \bibfield  {author} {\bibinfo {author} {\bibfnamefont {E.}~\bibnamefont
  {Breckenfeld}}, \bibinfo {author} {\bibfnamefont {N.}~\bibnamefont {Bronn}},
  \bibinfo {author} {\bibfnamefont {J.}~\bibnamefont {Karthik}}, \bibinfo
  {author} {\bibfnamefont {A.}~\bibnamefont {Damodaran}}, \bibinfo {author}
  {\bibfnamefont {S.}~\bibnamefont {Lee}}, \bibinfo {author} {\bibfnamefont
  {N.}~\bibnamefont {Mason}}, \ and\ \bibinfo {author} {\bibfnamefont
  {L.}~\bibnamefont {Martin}},\ }\href@noop {} {\bibfield  {journal} {\bibinfo
  {journal} {Phys. Rev. Lett.}\ }\textbf {\bibinfo {volume} {110}},\ \bibinfo
  {pages} {196804} (\bibinfo {year} {2013})}\BibitemShut {NoStop}%
\bibitem [{\citenamefont {Herranz}\ \emph {et~al.}(2007)\citenamefont
  {Herranz}, \citenamefont {Basleti{\'c}}, \citenamefont {Bibes}, \citenamefont
  {Carr{\'e}t{\'e}ro}, \citenamefont {Tafra}, \citenamefont {Jacquet},
  \citenamefont {Bouzehouane}, \citenamefont {Deranlot}, \citenamefont
  {Hamzi{\'c}}, \citenamefont {Broto},\ and\ \citenamefont
  {Barthélémy}}]{herranz2007high}%
  \BibitemOpen
  \bibfield  {author} {\bibinfo {author} {\bibfnamefont {G.}~\bibnamefont
  {Herranz}}, \bibinfo {author} {\bibfnamefont {M.}~\bibnamefont
  {Basleti{\'c}}}, \bibinfo {author} {\bibfnamefont {M.}~\bibnamefont {Bibes}},
  \bibinfo {author} {\bibfnamefont {C.}~\bibnamefont {Carr{\'e}t{\'e}ro}},
  \bibinfo {author} {\bibfnamefont {E.}~\bibnamefont {Tafra}}, \bibinfo
  {author} {\bibfnamefont {E.}~\bibnamefont {Jacquet}}, \bibinfo {author}
  {\bibfnamefont {K.}~\bibnamefont {Bouzehouane}}, \bibinfo {author}
  {\bibfnamefont {C.}~\bibnamefont {Deranlot}}, \bibinfo {author}
  {\bibfnamefont {A.}~\bibnamefont {Hamzi{\'c}}}, \bibinfo {author}
  {\bibfnamefont {J.}~\bibnamefont {Broto}}, \ and\ \bibinfo {author}
  {\bibfnamefont {A.}~\bibnamefont {Barthélémy}},\ }\href@noop {} {\bibfield
  {journal} {\bibinfo  {journal} {Phys. Rev. Lett.}\ }\textbf {\bibinfo
  {volume} {98}},\ \bibinfo {pages} {216803} (\bibinfo {year}
  {2007})}\BibitemShut {NoStop}%
\bibitem [{\citenamefont {Perna}\ \emph {et~al.}(2010)\citenamefont {Perna},
  \citenamefont {Maccariello}, \citenamefont {Radovic}, \citenamefont
  {Scotti~di Uccio}, \citenamefont {Pallecchi}, \citenamefont {Codda},
  \citenamefont {Marr{\'e}}, \citenamefont {Cantoni}, \citenamefont {Gazquez},
  \citenamefont {Varela} \emph {et~al.}}]{perna2010conducting}%
  \BibitemOpen
  \bibfield  {author} {\bibinfo {author} {\bibfnamefont {P.}~\bibnamefont
  {Perna}}, \bibinfo {author} {\bibfnamefont {D.}~\bibnamefont {Maccariello}},
  \bibinfo {author} {\bibfnamefont {M.}~\bibnamefont {Radovic}}, \bibinfo
  {author} {\bibfnamefont {U.}~\bibnamefont {Scotti~di Uccio}}, \bibinfo
  {author} {\bibfnamefont {I.}~\bibnamefont {Pallecchi}}, \bibinfo {author}
  {\bibfnamefont {M.}~\bibnamefont {Codda}}, \bibinfo {author} {\bibfnamefont
  {D.}~\bibnamefont {Marr{\'e}}}, \bibinfo {author} {\bibfnamefont
  {C.}~\bibnamefont {Cantoni}}, \bibinfo {author} {\bibfnamefont
  {J.}~\bibnamefont {Gazquez}}, \bibinfo {author} {\bibfnamefont
  {M.}~\bibnamefont {Varela}},  \emph {et~al.},\ }\href@noop {} {\bibfield
  {journal} {\bibinfo  {journal} {Appl. Phys. Lett.}\ }\textbf {\bibinfo
  {volume} {97}},\ \bibinfo {pages} {152111} (\bibinfo {year}
  {2010})}\BibitemShut {NoStop}%
\bibitem [{\citenamefont {Moetakef}\ \emph {et~al.}(2011)\citenamefont
  {Moetakef}, \citenamefont {Cain}, \citenamefont {Ouellette}, \citenamefont
  {Zhang}, \citenamefont {Klenov}, \citenamefont {Janotti}, \citenamefont
  {Van~de Walle}, \citenamefont {Rajan}, \citenamefont {Allen},\ and\
  \citenamefont {Stemmer}}]{moetakef2011electrostatic}%
  \BibitemOpen
  \bibfield  {author} {\bibinfo {author} {\bibfnamefont {P.}~\bibnamefont
  {Moetakef}}, \bibinfo {author} {\bibfnamefont {T.~A.}\ \bibnamefont {Cain}},
  \bibinfo {author} {\bibfnamefont {D.~G.}\ \bibnamefont {Ouellette}}, \bibinfo
  {author} {\bibfnamefont {J.~Y.}\ \bibnamefont {Zhang}}, \bibinfo {author}
  {\bibfnamefont {D.~O.}\ \bibnamefont {Klenov}}, \bibinfo {author}
  {\bibfnamefont {A.}~\bibnamefont {Janotti}}, \bibinfo {author} {\bibfnamefont
  {C.~G.}\ \bibnamefont {Van~de Walle}}, \bibinfo {author} {\bibfnamefont
  {S.}~\bibnamefont {Rajan}}, \bibinfo {author} {\bibfnamefont {S.~J.}\
  \bibnamefont {Allen}}, \ and\ \bibinfo {author} {\bibfnamefont
  {S.}~\bibnamefont {Stemmer}},\ }\href@noop {} {\bibfield  {journal} {\bibinfo
   {journal} {Appl. Phys. Lett.}\ }\textbf {\bibinfo {volume} {99}},\ \bibinfo
  {pages} {232116} (\bibinfo {year} {2011})}\BibitemShut {NoStop}%
\bibitem [{\citenamefont {Kornblum}\ \emph {et~al.}(2015)\citenamefont
  {Kornblum}, \citenamefont {Jin}, \citenamefont {Kumah}, \citenamefont
  {Ernst}, \citenamefont {Broadbridge}, \citenamefont {Ahn},\ and\
  \citenamefont {Walker}}]{kornblum2015oxide}%
  \BibitemOpen
  \bibfield  {author} {\bibinfo {author} {\bibfnamefont {L.}~\bibnamefont
  {Kornblum}}, \bibinfo {author} {\bibfnamefont {E.~N.}\ \bibnamefont {Jin}},
  \bibinfo {author} {\bibfnamefont {D.~P.}\ \bibnamefont {Kumah}}, \bibinfo
  {author} {\bibfnamefont {A.~T.}\ \bibnamefont {Ernst}}, \bibinfo {author}
  {\bibfnamefont {C.~C.}\ \bibnamefont {Broadbridge}}, \bibinfo {author}
  {\bibfnamefont {C.~H.}\ \bibnamefont {Ahn}}, \ and\ \bibinfo {author}
  {\bibfnamefont {F.~J.}\ \bibnamefont {Walker}},\ }\href@noop {} {\bibfield
  {journal} {\bibinfo  {journal} {Appl. Phys. Lett.}\ }\textbf {\bibinfo
  {volume} {106}},\ \bibinfo {pages} {201602} (\bibinfo {year}
  {2015})}\BibitemShut {NoStop}%
\bibitem [{\citenamefont {Marshall}\ \emph {et~al.}(2016)\citenamefont
  {Marshall}, \citenamefont {Mikheev}, \citenamefont {Raghavan},\ and\
  \citenamefont {Stemmer}}]{marshall2016pseudogaps}%
  \BibitemOpen
  \bibfield  {author} {\bibinfo {author} {\bibfnamefont {P.~B.}\ \bibnamefont
  {Marshall}}, \bibinfo {author} {\bibfnamefont {E.}~\bibnamefont {Mikheev}},
  \bibinfo {author} {\bibfnamefont {S.}~\bibnamefont {Raghavan}}, \ and\
  \bibinfo {author} {\bibfnamefont {S.}~\bibnamefont {Stemmer}},\ }\href@noop
  {} {\bibfield  {journal} {\bibinfo  {journal} {Phys. Rev. Lett.}\ }\textbf
  {\bibinfo {volume} {117}},\ \bibinfo {pages} {046402} (\bibinfo {year}
  {2016})}\BibitemShut {NoStop}%
\bibitem [{\citenamefont {He}\ \emph {et~al.}(2012)\citenamefont {He},
  \citenamefont {Sanders}, \citenamefont {Gray}, \citenamefont {Wong},
  \citenamefont {Mehta},\ and\ \citenamefont {Suzuki}}]{he2012metal}%
  \BibitemOpen
  \bibfield  {author} {\bibinfo {author} {\bibfnamefont {C.}~\bibnamefont
  {He}}, \bibinfo {author} {\bibfnamefont {T.}~\bibnamefont {Sanders}},
  \bibinfo {author} {\bibfnamefont {M.}~\bibnamefont {Gray}}, \bibinfo {author}
  {\bibfnamefont {F.}~\bibnamefont {Wong}}, \bibinfo {author} {\bibfnamefont
  {V.}~\bibnamefont {Mehta}}, \ and\ \bibinfo {author} {\bibfnamefont
  {Y.}~\bibnamefont {Suzuki}},\ }\href@noop {} {\bibfield  {journal} {\bibinfo
  {journal} {Phys. Rev. B}\ }\textbf {\bibinfo {volume} {86}},\ \bibinfo
  {pages} {081401} (\bibinfo {year} {2012})}\BibitemShut {NoStop}%
\bibitem [{\citenamefont {Ahmadi-Majlan}\ \emph {et~al.}(2018)\citenamefont
  {Ahmadi-Majlan}, \citenamefont {Chen}, \citenamefont {Lim}, \citenamefont
  {Conlin}, \citenamefont {Hensley}, \citenamefont {Chrysler}, \citenamefont
  {Su}, \citenamefont {Chen}, \citenamefont {Kumah},\ and\ \citenamefont
  {Ngai}}]{ahmadi2018tuning}%
  \BibitemOpen
  \bibfield  {author} {\bibinfo {author} {\bibfnamefont {K.}~\bibnamefont
  {Ahmadi-Majlan}}, \bibinfo {author} {\bibfnamefont {T.}~\bibnamefont {Chen}},
  \bibinfo {author} {\bibfnamefont {Z.~H.}\ \bibnamefont {Lim}}, \bibinfo
  {author} {\bibfnamefont {P.}~\bibnamefont {Conlin}}, \bibinfo {author}
  {\bibfnamefont {R.}~\bibnamefont {Hensley}}, \bibinfo {author} {\bibfnamefont
  {M.}~\bibnamefont {Chrysler}}, \bibinfo {author} {\bibfnamefont
  {D.}~\bibnamefont {Su}}, \bibinfo {author} {\bibfnamefont {H.}~\bibnamefont
  {Chen}}, \bibinfo {author} {\bibfnamefont {D.~P.}\ \bibnamefont {Kumah}}, \
  and\ \bibinfo {author} {\bibfnamefont {J.~H.}\ \bibnamefont {Ngai}},\
  }\href@noop {} {\bibfield  {journal} {\bibinfo  {journal} {Appl. Phys.
  Lett.}\ }\textbf {\bibinfo {volume} {112}},\ \bibinfo {pages} {193104}
  (\bibinfo {year} {2018})}\BibitemShut {NoStop}%
\bibitem [{\citenamefont {Chambers}\ \emph {et~al.}(2011)\citenamefont
  {Chambers}, \citenamefont {Qiao}, \citenamefont {Droubay}, \citenamefont
  {Kaspar}, \citenamefont {Arey},\ and\ \citenamefont
  {Sushko}}]{chambers2011band}%
  \BibitemOpen
  \bibfield  {author} {\bibinfo {author} {\bibfnamefont {S.~A.}\ \bibnamefont
  {Chambers}}, \bibinfo {author} {\bibfnamefont {L.}~\bibnamefont {Qiao}},
  \bibinfo {author} {\bibfnamefont {T.~C.}\ \bibnamefont {Droubay}}, \bibinfo
  {author} {\bibfnamefont {T.~C.}\ \bibnamefont {Kaspar}}, \bibinfo {author}
  {\bibfnamefont {B.~W.}\ \bibnamefont {Arey}}, \ and\ \bibinfo {author}
  {\bibfnamefont {P.}~\bibnamefont {Sushko}},\ }\href@noop {} {\bibfield
  {journal} {\bibinfo  {journal} {Phys. Rev. Lett.}\ }\textbf {\bibinfo
  {volume} {107}},\ \bibinfo {pages} {206802} (\bibinfo {year}
  {2011})}\BibitemShut {NoStop}%
\bibitem [{\citenamefont {Comes}\ \emph {et~al.}(2017)\citenamefont {Comes},
  \citenamefont {Spurgeon}, \citenamefont {Kepaptsoglou}, \citenamefont
  {Engelhard}, \citenamefont {Perea}, \citenamefont {Kaspar}, \citenamefont
  {Ramasse}, \citenamefont {Sushko},\ and\ \citenamefont
  {Chambers}}]{comes2017probing}%
  \BibitemOpen
  \bibfield  {author} {\bibinfo {author} {\bibfnamefont {R.~B.}\ \bibnamefont
  {Comes}}, \bibinfo {author} {\bibfnamefont {S.~R.}\ \bibnamefont {Spurgeon}},
  \bibinfo {author} {\bibfnamefont {D.~M.}\ \bibnamefont {Kepaptsoglou}},
  \bibinfo {author} {\bibfnamefont {M.~H.}\ \bibnamefont {Engelhard}}, \bibinfo
  {author} {\bibfnamefont {D.~E.}\ \bibnamefont {Perea}}, \bibinfo {author}
  {\bibfnamefont {T.~C.}\ \bibnamefont {Kaspar}}, \bibinfo {author}
  {\bibfnamefont {Q.~M.}\ \bibnamefont {Ramasse}}, \bibinfo {author}
  {\bibfnamefont {P.~V.}\ \bibnamefont {Sushko}}, \ and\ \bibinfo {author}
  {\bibfnamefont {S.~A.}\ \bibnamefont {Chambers}},\ }\href@noop {} {\bibfield
  {journal} {\bibinfo  {journal} {Chem. Mater.}\ }\textbf {\bibinfo {volume}
  {29}},\ \bibinfo {pages} {1147} (\bibinfo {year} {2017})}\BibitemShut
  {NoStop}%
\bibitem [{\citenamefont {F{\^e}te}\ \emph {et~al.}(2015)\citenamefont
  {F{\^e}te}, \citenamefont {Cancellieri}, \citenamefont {Li}, \citenamefont
  {Stornaiuolo}, \citenamefont {Caviglia}, \citenamefont {Gariglio},\ and\
  \citenamefont {Triscone}}]{fete2015growth}%
  \BibitemOpen
  \bibfield  {author} {\bibinfo {author} {\bibfnamefont {A.}~\bibnamefont
  {F{\^e}te}}, \bibinfo {author} {\bibfnamefont {C.}~\bibnamefont
  {Cancellieri}}, \bibinfo {author} {\bibfnamefont {D.}~\bibnamefont {Li}},
  \bibinfo {author} {\bibfnamefont {D.}~\bibnamefont {Stornaiuolo}}, \bibinfo
  {author} {\bibfnamefont {A.}~\bibnamefont {Caviglia}}, \bibinfo {author}
  {\bibfnamefont {S.}~\bibnamefont {Gariglio}}, \ and\ \bibinfo {author}
  {\bibfnamefont {J.-M.}\ \bibnamefont {Triscone}},\ }\href@noop {} {\bibfield
  {journal} {\bibinfo  {journal} {Applied Physics Letters}\ }\textbf {\bibinfo
  {volume} {106}},\ \bibinfo {pages} {051604} (\bibinfo {year}
  {2015})}\BibitemShut {NoStop}%
\bibitem [{\citenamefont {Arima}\ \emph {et~al.}(1993)\citenamefont {Arima},
  \citenamefont {Tokura},\ and\ \citenamefont {Torrance}}]{arima1993variation}%
  \BibitemOpen
  \bibfield  {author} {\bibinfo {author} {\bibfnamefont {T.}~\bibnamefont
  {Arima}}, \bibinfo {author} {\bibfnamefont {Y.}~\bibnamefont {Tokura}}, \
  and\ \bibinfo {author} {\bibfnamefont {J.}~\bibnamefont {Torrance}},\
  }\href@noop {} {\bibfield  {journal} {\bibinfo  {journal} {Phys. Rev. B}\
  }\textbf {\bibinfo {volume} {48}},\ \bibinfo {pages} {17006} (\bibinfo {year}
  {1993})}\BibitemShut {NoStop}%
\bibitem [{\citenamefont {Takuya}\ \emph {et~al.}(2000)\citenamefont {Takuya},
  \citenamefont {Naoto}, \citenamefont {Akira}, \citenamefont {Kazunari},
  \citenamefont {Tsuda}, \citenamefont {Michiyoshi}, \citenamefont {Kenichi},
  \citenamefont {Takashi}, \citenamefont {Kaori}, \citenamefont {Hiroaki},\
  and\ \citenamefont {Masayuki}}]{HASHIMOTO2000181}%
  \BibitemOpen
  \bibfield  {author} {\bibinfo {author} {\bibfnamefont {H.}~\bibnamefont
  {Takuya}}, \bibinfo {author} {\bibfnamefont {T.}~\bibnamefont {Naoto}},
  \bibinfo {author} {\bibfnamefont {K.}~\bibnamefont {Akira}}, \bibinfo
  {author} {\bibfnamefont {T.}~\bibnamefont {Kazunari}}, \bibinfo {author}
  {\bibfnamefont {K.}~\bibnamefont {Tsuda}}, \bibinfo {author} {\bibfnamefont
  {T.}~\bibnamefont {Michiyoshi}}, \bibinfo {author} {\bibfnamefont
  {O.}~\bibnamefont {Kenichi}}, \bibinfo {author} {\bibfnamefont
  {K.}~\bibnamefont {Takashi}}, \bibinfo {author} {\bibfnamefont
  {Y.}~\bibnamefont {Kaori}}, \bibinfo {author} {\bibfnamefont
  {T.}~\bibnamefont {Hiroaki}}, \ and\ \bibinfo {author} {\bibfnamefont
  {D.}~\bibnamefont {Masayuki}},\ }\href {\doibase
  https://doi.org/10.1016/S0167-2738(00)00657-3} {\bibfield  {journal}
  {\bibinfo  {journal} {Solid State Ion.}\ }\textbf {\bibinfo {volume} {132}},\
  \bibinfo {pages} {181 } (\bibinfo {year} {2000})},\ \bibinfo {note} {solid
  Oxide Fuel Cells dedicated to Prof. H. Tagawa}\BibitemShut {NoStop}%
\bibitem [{\citenamefont {Kalabukhov}\ \emph {et~al.}(2007)\citenamefont
  {Kalabukhov}, \citenamefont {Gunnarsson}, \citenamefont {B{\"o}rjesson},
  \citenamefont {Olsson}, \citenamefont {Claeson},\ and\ \citenamefont
  {Winkler}}]{kalabukhov2007effect}%
  \BibitemOpen
  \bibfield  {author} {\bibinfo {author} {\bibfnamefont {A.}~\bibnamefont
  {Kalabukhov}}, \bibinfo {author} {\bibfnamefont {R.}~\bibnamefont
  {Gunnarsson}}, \bibinfo {author} {\bibfnamefont {J.}~\bibnamefont
  {B{\"o}rjesson}}, \bibinfo {author} {\bibfnamefont {E.}~\bibnamefont
  {Olsson}}, \bibinfo {author} {\bibfnamefont {T.}~\bibnamefont {Claeson}}, \
  and\ \bibinfo {author} {\bibfnamefont {D.}~\bibnamefont {Winkler}},\
  }\href@noop {} {\bibfield  {journal} {\bibinfo  {journal} {Phys. Rev. B}\
  }\textbf {\bibinfo {volume} {75}},\ \bibinfo {pages} {121404} (\bibinfo
  {year} {2007})}\BibitemShut {NoStop}%
\bibitem [{\citenamefont {Kawasaki}\ \emph {et~al.}(1994)\citenamefont
  {Kawasaki}, \citenamefont {Takahashi}, \citenamefont {Maeda}, \citenamefont
  {Tsuchiya}, \citenamefont {Shinohara}, \citenamefont {Ishiyama},
  \citenamefont {Yonezawa}, \citenamefont {Yoshimoto},\ and\ \citenamefont
  {Koinuma}}]{kawasaki1994atomic}%
  \BibitemOpen
  \bibfield  {author} {\bibinfo {author} {\bibfnamefont {M.}~\bibnamefont
  {Kawasaki}}, \bibinfo {author} {\bibfnamefont {K.}~\bibnamefont {Takahashi}},
  \bibinfo {author} {\bibfnamefont {T.}~\bibnamefont {Maeda}}, \bibinfo
  {author} {\bibfnamefont {R.}~\bibnamefont {Tsuchiya}}, \bibinfo {author}
  {\bibfnamefont {M.}~\bibnamefont {Shinohara}}, \bibinfo {author}
  {\bibfnamefont {O.}~\bibnamefont {Ishiyama}}, \bibinfo {author}
  {\bibfnamefont {T.}~\bibnamefont {Yonezawa}}, \bibinfo {author}
  {\bibfnamefont {M.}~\bibnamefont {Yoshimoto}}, \ and\ \bibinfo {author}
  {\bibfnamefont {H.}~\bibnamefont {Koinuma}},\ }\href@noop {} {\bibfield
  {journal} {\bibinfo  {journal} {Science}\ }\textbf {\bibinfo {volume}
  {266}},\ \bibinfo {pages} {1540} (\bibinfo {year} {1994})}\BibitemShut
  {NoStop}%
\bibitem [{\citenamefont {Bj{\"o}rck}\ and\ \citenamefont
  {Andersson}(2007)}]{bjorck2007genx}%
  \BibitemOpen
  \bibfield  {author} {\bibinfo {author} {\bibfnamefont {M.}~\bibnamefont
  {Bj{\"o}rck}}\ and\ \bibinfo {author} {\bibfnamefont {G.}~\bibnamefont
  {Andersson}},\ }\href@noop {} {\bibfield  {journal} {\bibinfo  {journal} {J.
  Appl. Crystallogr.}\ }\textbf {\bibinfo {volume} {40}},\ \bibinfo {pages}
  {1174} (\bibinfo {year} {2007})}\BibitemShut {NoStop}%
\bibitem [{\citenamefont {Koohfar}\ \emph {et~al.}(2017)\citenamefont
  {Koohfar}, \citenamefont {Disa}, \citenamefont {Marshall}, \citenamefont
  {Walker}, \citenamefont {Ahn},\ and\ \citenamefont
  {Kumah}}]{koohfar2017structural}%
  \BibitemOpen
  \bibfield  {author} {\bibinfo {author} {\bibfnamefont {S.}~\bibnamefont
  {Koohfar}}, \bibinfo {author} {\bibfnamefont {A.}~\bibnamefont {Disa}},
  \bibinfo {author} {\bibfnamefont {M.}~\bibnamefont {Marshall}}, \bibinfo
  {author} {\bibfnamefont {F.}~\bibnamefont {Walker}}, \bibinfo {author}
  {\bibfnamefont {C.}~\bibnamefont {Ahn}}, \ and\ \bibinfo {author}
  {\bibfnamefont {D.}~\bibnamefont {Kumah}},\ }\href@noop {} {\bibfield
  {journal} {\bibinfo  {journal} {Phys. Rev. B}\ }\textbf {\bibinfo {volume}
  {96}},\ \bibinfo {pages} {024108} (\bibinfo {year} {2017})}\BibitemShut
  {NoStop}%
\bibitem [{\citenamefont {Maiti}\ and\ \citenamefont
  {Sarma}(1996)}]{maiti1996electronic}%
  \BibitemOpen
  \bibfield  {author} {\bibinfo {author} {\bibfnamefont {K.}~\bibnamefont
  {Maiti}}\ and\ \bibinfo {author} {\bibfnamefont {D.}~\bibnamefont {Sarma}},\
  }\href@noop {} {\bibfield  {journal} {\bibinfo  {journal} {Phys. Rev. B}\
  }\textbf {\bibinfo {volume} {54}},\ \bibinfo {pages} {7816} (\bibinfo {year}
  {1996})}\BibitemShut {NoStop}%
\bibitem [{\citenamefont {Colby}\ \emph {et~al.}(2013)\citenamefont {Colby},
  \citenamefont {Qiao}, \citenamefont {Zhang}, \citenamefont {Shutthanandan},
  \citenamefont {Ciston}, \citenamefont {Kabius},\ and\ \citenamefont
  {Chambers}}]{colby2013cation}%
  \BibitemOpen
  \bibfield  {author} {\bibinfo {author} {\bibfnamefont {R.}~\bibnamefont
  {Colby}}, \bibinfo {author} {\bibfnamefont {L.}~\bibnamefont {Qiao}},
  \bibinfo {author} {\bibfnamefont {K.}~\bibnamefont {Zhang}}, \bibinfo
  {author} {\bibfnamefont {V.}~\bibnamefont {Shutthanandan}}, \bibinfo {author}
  {\bibfnamefont {J.}~\bibnamefont {Ciston}}, \bibinfo {author} {\bibfnamefont
  {B.}~\bibnamefont {Kabius}}, \ and\ \bibinfo {author} {\bibfnamefont {S.~A.}\
  \bibnamefont {Chambers}},\ }\href@noop {} {\bibfield  {journal} {\bibinfo
  {journal} {Phys. Rev. B}\ }\textbf {\bibinfo {volume} {88}},\ \bibinfo
  {pages} {155325} (\bibinfo {year} {2013})}\BibitemShut {NoStop}%
\bibitem [{\citenamefont {Qiao}\ \emph {et~al.}(2011)\citenamefont {Qiao},
  \citenamefont {Droubay}, \citenamefont {Bowden}, \citenamefont
  {Shutthanandan}, \citenamefont {Kaspar},\ and\ \citenamefont
  {Chambers}}]{qiao2011lacro3}%
  \BibitemOpen
  \bibfield  {author} {\bibinfo {author} {\bibfnamefont {L.}~\bibnamefont
  {Qiao}}, \bibinfo {author} {\bibfnamefont {T.~C.}\ \bibnamefont {Droubay}},
  \bibinfo {author} {\bibfnamefont {M.~E.}\ \bibnamefont {Bowden}}, \bibinfo
  {author} {\bibfnamefont {V.}~\bibnamefont {Shutthanandan}}, \bibinfo {author}
  {\bibfnamefont {T.~C.}\ \bibnamefont {Kaspar}}, \ and\ \bibinfo {author}
  {\bibfnamefont {S.~A.}\ \bibnamefont {Chambers}},\ }\href@noop {} {\bibfield
  {journal} {\bibinfo  {journal} {Appl. Phys. Lett.}\ }\textbf {\bibinfo
  {volume} {99}},\ \bibinfo {pages} {061904} (\bibinfo {year}
  {2011})}\BibitemShut {NoStop}%
\bibitem [{\citenamefont {Zhang}\ \emph {et~al.}(2015)\citenamefont {Zhang},
  \citenamefont {Du}, \citenamefont {Sushko}, \citenamefont {Bowden},
  \citenamefont {Shutthanandan}, \citenamefont {Sallis}, \citenamefont
  {Piper},\ and\ \citenamefont {Chambers}}]{zhang2015hole}%
  \BibitemOpen
  \bibfield  {author} {\bibinfo {author} {\bibfnamefont {K.}~\bibnamefont
  {Zhang}}, \bibinfo {author} {\bibfnamefont {Y.}~\bibnamefont {Du}}, \bibinfo
  {author} {\bibfnamefont {P.}~\bibnamefont {Sushko}}, \bibinfo {author}
  {\bibfnamefont {M.~E.}\ \bibnamefont {Bowden}}, \bibinfo {author}
  {\bibfnamefont {V.}~\bibnamefont {Shutthanandan}}, \bibinfo {author}
  {\bibfnamefont {S.}~\bibnamefont {Sallis}}, \bibinfo {author} {\bibfnamefont
  {L.~F.}\ \bibnamefont {Piper}}, \ and\ \bibinfo {author} {\bibfnamefont
  {S.~A.}\ \bibnamefont {Chambers}},\ }\href@noop {} {\bibfield  {journal}
  {\bibinfo  {journal} {Phys. Rev. B}\ }\textbf {\bibinfo {volume} {91}},\
  \bibinfo {pages} {155129} (\bibinfo {year} {2015})}\BibitemShut {NoStop}%
\bibitem [{\citenamefont {Plumb}\ \emph {et~al.}(2014)\citenamefont {Plumb},
  \citenamefont {Salluzzo}, \citenamefont {Razzoli}, \citenamefont
  {M{\aa}nsson}, \citenamefont {Falub}, \citenamefont {Krempasky},
  \citenamefont {Matt}, \citenamefont {Chang}, \citenamefont {Schulte},
  \citenamefont {Braun} \emph {et~al.}}]{plumb2014mixed}%
  \BibitemOpen
  \bibfield  {author} {\bibinfo {author} {\bibfnamefont {N.}~\bibnamefont
  {Plumb}}, \bibinfo {author} {\bibfnamefont {M.}~\bibnamefont {Salluzzo}},
  \bibinfo {author} {\bibfnamefont {E.}~\bibnamefont {Razzoli}}, \bibinfo
  {author} {\bibfnamefont {M.}~\bibnamefont {M{\aa}nsson}}, \bibinfo {author}
  {\bibfnamefont {M.}~\bibnamefont {Falub}}, \bibinfo {author} {\bibfnamefont
  {J.}~\bibnamefont {Krempasky}}, \bibinfo {author} {\bibfnamefont
  {C.}~\bibnamefont {Matt}}, \bibinfo {author} {\bibfnamefont {J.}~\bibnamefont
  {Chang}}, \bibinfo {author} {\bibfnamefont {M.}~\bibnamefont {Schulte}},
  \bibinfo {author} {\bibfnamefont {J.}~\bibnamefont {Braun}},  \emph
  {et~al.},\ }\href@noop {} {\bibfield  {journal} {\bibinfo  {journal} {Phys.
  Rev. Lett.}\ }\textbf {\bibinfo {volume} {113}},\ \bibinfo {pages} {086801}
  (\bibinfo {year} {2014})}\BibitemShut {NoStop}%
\bibitem [{\citenamefont {Weinberg}\ and\ \citenamefont
  {Larssen}(1961)}]{weinberg1961electron}%
  \BibitemOpen
  \bibfield  {author} {\bibinfo {author} {\bibfnamefont {I.}~\bibnamefont
  {Weinberg}}\ and\ \bibinfo {author} {\bibfnamefont {P.}~\bibnamefont
  {Larssen}},\ }\href@noop {} {\bibfield  {journal} {\bibinfo  {journal}
  {Nature}\ }\textbf {\bibinfo {volume} {192}},\ \bibinfo {pages} {445}
  (\bibinfo {year} {1961})}\BibitemShut {NoStop}%
\bibitem [{\citenamefont {Ueda}\ \emph {et~al.}(1998)\citenamefont {Ueda},
  \citenamefont {Tabata},\ and\ \citenamefont
  {Kawai}}]{ueda1998ferromagnetism}%
  \BibitemOpen
  \bibfield  {author} {\bibinfo {author} {\bibfnamefont {K.}~\bibnamefont
  {Ueda}}, \bibinfo {author} {\bibfnamefont {H.}~\bibnamefont {Tabata}}, \ and\
  \bibinfo {author} {\bibfnamefont {T.}~\bibnamefont {Kawai}},\ }\href@noop {}
  {\bibfield  {journal} {\bibinfo  {journal} {Science}\ }\textbf {\bibinfo
  {volume} {280}},\ \bibinfo {pages} {1064} (\bibinfo {year}
  {1998})}\BibitemShut {NoStop}%
\bibitem [{\citenamefont {Zhou}\ \emph {et~al.}(2011)\citenamefont {Zhou},
  \citenamefont {Alonso}, \citenamefont {Muonz}, \citenamefont
  {Fern{\'a}ndez-D{\'\i}az},\ and\ \citenamefont
  {Goodenough}}]{zhou2011magnetic}%
  \BibitemOpen
  \bibfield  {author} {\bibinfo {author} {\bibfnamefont {J.-S.}\ \bibnamefont
  {Zhou}}, \bibinfo {author} {\bibfnamefont {J.}~\bibnamefont {Alonso}},
  \bibinfo {author} {\bibfnamefont {A.}~\bibnamefont {Muonz}}, \bibinfo
  {author} {\bibfnamefont {M.}~\bibnamefont {Fern{\'a}ndez-D{\'\i}az}}, \ and\
  \bibinfo {author} {\bibfnamefont {J.}~\bibnamefont {Goodenough}},\
  }\href@noop {} {\bibfield  {journal} {\bibinfo  {journal} {Phys. Rev. Lett.}\
  }\textbf {\bibinfo {volume} {106}},\ \bibinfo {pages} {057201} (\bibinfo
  {year} {2011})}\BibitemShut {NoStop}%
\end{thebibliography}
%

\end{document}